\documentstyle[11pt,moriond,epsfig]{article}

\def\Journal#1#2#3#4{{#1} {\bf #2}, #3 (#4)}

\def\PLB{{\em Phys. Lett.}  B}
\def\PRL{\em Phys. Rev. Lett.}
\def\PRD{{\em Phys. Rev.} D}

\def\be{\begin{equation}}
\def\ee{\end{equation}}
\def\bea{\begin{eqnarray}}
\def\eea{\end{eqnarray}}

\begin{document}
\vspace*{4cm}
\title{SEARCH FOR STIMULATED NEUTRINO CONVERSION WITH AN RF CAVITY
\footnote{Contributed paper to the ``Rencontres de Blois'', Blois,
France, June 28-July 3 1999}}
\author{F. Vannucci}

\address{LPNHE Universit{\'e} Paris 7,\\
Paris, France}

\maketitle\abstracts{
Little is known on radiative decays of $\nu_{\mu}$ and  $\nu_{\tau}$. Lifetimes can be constrained, or the process discovered, by a search for flavour conversion in a superconducting RF cavity installed in a neutrino beam. The lifetimes tested with such a method can be of interest for astrophysics and cosmology. In particular the conjecture which explains both the LSND and the SuperKamiokande signals by radiative decays of  $\nu_{\mu}$ can be straightforwardly put to the test.}

\section{INTRODUCTION}

Neutrino decays require flavour violation and the existence of massive neutrino species. After the recent announcement of an evidence for oscillations presented by the SuperKamiokande collaboration \cite{fu}, these two points seem  fulfilled.

Radiative decays with the emission of one photon represent the most plausible mode between neutrinos having masses in the range of a few eV or less. Radiative decays in vacuum have been calculated long ago \cite{pa}. The process is GIM suppressed and has an immeasurably long lifetime for masses of present interest.

 Few experimental limits exist. One comes from the SN1987A burst and applies to  $\bar{\nu_{e}}$ since only  $\bar{\nu_{e}}$ have been detected in that circumstance. The limit is \cite{ch}:

  $\tau_{0}/m > 6\: 10^{15}$ s/eV

A limit on  $\nu_{\mu}$ decay comes from a Los Alamos measurement \cite{kr}: 

  $\tau_{0}/m > 15.4$ s/eV

Both limits apply to decays between neutrinos having a strong mass hierarchy. For mass-degenerate neutrinos, two results have been published. The first one obtained at a nuclear reactor applies to ${\bar\nu_{e}}$ \cite{bo}. The second one has been set with solar neutrinos during a total eclipse \cite{bi}. All these limits are very far from the theoretical expectations.

It has been recently recognised that the radiative decay probability is very much amplified in dense media. For example matter effects can hugely stimulate the process \cite{do}. The amplification factor relates the lifetime in vacuum $\tau_{0}$ to the one in matter $\tau_{m}$ by the formula:

$ \tau_{0}/\tau_{m} = 8.6\: 10^{23} F(v)(N_e/10^{24}cm^{-3})^2(1eV/m)^4(m^2/\delta m^2)^2 $ 

where $N_e$ is the electron density of the medium, m is the mass of the decaying neutrino and $\delta m^2$ the difference between the squared masses of the participating particles (the same $\delta m^2$ as the one appearing in oscillatiom searches). For relativistic neutrinos F(v) takes the value 4m/E, E being the neutrino energy. The formula applies to the case of mass-degenerate neutrinos. For the case of mass hierarchy, one should take $\delta m^2 = m^2$.

Enhanced decays in media could have important consequences for astrophysical considerations. In particular, it has been conjectured that the LSND appearance of ${\bar\nu_{e}}$, and the SuperKamiokande disappearance of $\nu_{\mu}$ could have the same origin in the radiative decay of  $\nu_{\mu}$ amplified in matter \cite{va}. The process in question is:

$\nu_{\mu} \rightarrow \gamma + {\bar\nu_{e}}$
within a scheme of mass-degenerate neutrinos.

The corresponding lifetime in matter is found to be :

$\tau_{m} = 4\: 10^{-13}$ s/eV.

This seems extremely small, but with the amplification factor due to the presence of matter, it corresponds to the equivalent lifetime in vacuum:

$\tau_{0} = 9.5\: 10^4 m^2 /(\delta m^2)^2$ 

Thus $\tau_{0}$ can be very large if $\delta m^2$ is small. For example with the value $10^{-5} eV^2$ favoured by the MSW solution of the solar deficit, and with m of the order of a few eV as required by cosmology, the lifetime reaches a value close to the age of the universe.

\section{RF STIMULATED CONVERSION}
An efficient method to stimulate conversions between neutrino flavours consists in sending a high energy neutrino beam through a strong electromagnetic field. The case of a high quality RF cavity has been considered and the probability of conversion calculated in Ref \cite{ma}. We show how this idea can set interesting limits.

In a high energy  $\nu_{\mu}$ beam, a high quality RF cavity gives a target of very low energy photons which can induce the conversion:

$\nu_{\mu}\rightarrow{\bar\nu_{e}}$     or  $\nu_{\mu}\rightarrow{\bar\nu_{\tau}}$   

Experimentally, if neutrinos are Majorana particles, the first process shows as an excess of interactions producing a positron. If neutrinos are Dirac particles, or if a  $\nu_{\tau}$ is produced, the process results in a disappearance of CC  $\nu_{\mu}$ interactions. The transition rate depends linearly on the quality factor Q of the cavity and on the power P stored into it. To optimize the search, the cavity must have the highest possible Q*P and a large surface area intersecting the beam in order to maximize the number of crossing neutrinos. A superconducting cavity is required.

In the fundamental mode $TM_{010}$, the diameter of a cavity is inversely proportional to its frequency. This favours a frequency in the range 300-500 Mhz for which Q is limited by the residual resistance and does not necessitate the lowest temperature of superfluid helium.
Note that the cavity is used in a continuous mode. This is very different from the pulsed mode compulsory to accelerate beams. The power supplied to the cavity is very limited and only necessary to compensate for the losses in the superconductor. The stored energy is proportional to $V_{c}^2$, where $V_{c}$ is the accelerating tension.

A superconducting cavity developped for the French synchrotron radiation project Soleil could be available for the present search.  A good working point is obtained for $V_{c}=6\: MV/m$ for which $Q = 3\: 10^9$ and the energy is calculated to be 33 J.  For the neutrino beam, we will consider the case of the K2K beam built at KEK \cite{oy}. It aims at the SuperKamiokande detector and is devoted to confirming the signal of oscillations found with atmospheric neutrinos. On the KEK site the beam is essentially composed of  $\nu_{\mu}$ with a peak energy around 1.4 GeV. The intensity reaches $1.5\: 10^5$ neutrinos /$cm^2$ /spill at a distance of 300 m from the production target where two front detectors have been built.

The cavity covers a surface area of $0.45\: m^2$. The neutrino flux crossing this area results in more than 100 CC  $\nu_{\mu}$ events per day or 3000 during a one month exposure time in the existing front detectors. This also produces  30 CC  $\nu_{e}$ events from normal beam contamination.

\section{ACHIEVABLE LIMITS}

The proposed test consists in comparing the flux of neutrinos of various flavours with and without the cavity switched on. The experiment measures the number of CC  $\nu_{\mu}$ interactions and the ratio electron/muon in the detectors with the cavity on and off, and looks for a difference. A $2\%$ conversion seems accesible for the  $\nu_{e}$ appearance mode. This would result in 60 excess electron (positron) events above the expected 30. For the disappearance channel a $5\%$ limit is assumed. This level corresponds to a statistical effect of 3$\sigma$. With these limits it is possible to extract a lower limit on the radiative lifetimes as shown in Figure 1.

\begin{figure}[h]
\centerline{\epsfig{file=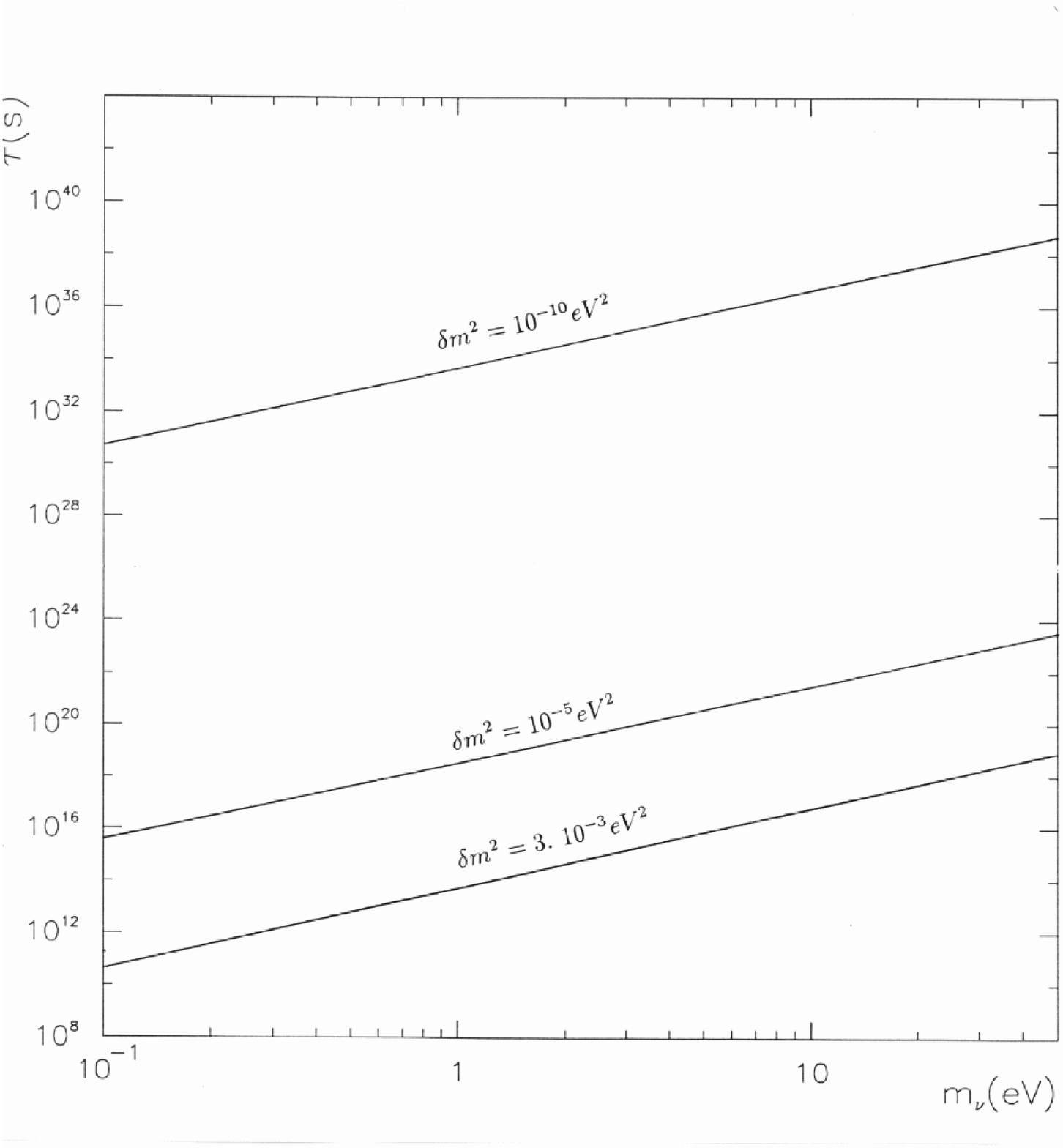,width=0.6\linewidth}}
\end{figure}
 The curves labelled $10^{-5}$ and $10^{-10}$ assume the $\delta m^2$ favoured by the deficit of solar neutrinos (MSW and just-so solutions respectively). They relate to the mode:

 $\nu_{\mu}\rightarrow\gamma +\nu_{e}$

Very interesting limits can be reached, for example:

$\tau_{0} > 5\: 10^{18}$ s     for      $\delta m^2= 10^{-5} eV^2$     and   m = 1 eV

The curve corresponding to $\delta m^2= 3\: 10^{-3}$ assumes the value of $\delta m^2$ given by the atmospheric neutrino deficit and applies to the mode:

 $\nu_{\tau}\rightarrow\gamma +\nu_{\mu}$

These curves are obtained for mass-degenerate neutrinos in the range of masses interesting for cosmology (1-10 eV).
One can also extract a limit for the case of a strong mass hierarchy. From the signal of atmospheric neutrinos, one infers a  $\nu_{\tau}$ mass of 50 meV. For such a mass the lifetime limit is:

$\tau_{0} > 8\:10^7$ s

The limit becomes:

$\tau_{0} > 5\:10^3$ s

for a mass of 1 eV.
These limits apply to both  $\nu_{\mu}$ and  $\nu_{\tau}$. For  $\nu_{\mu}$, this is more than two orders of magnitude better than the present experimental limit. For  $\nu_{\tau}$, this would constitute the first laboratory limit.

The measured conversion rate in the cavity R is related to the lifetime in vacuum by the formula:

$\tau_{0} = 100/R\:(m/\delta m^2)^3 $

With the lifetime found in interpreting the LSND and SuperKamiokande results \cite{va}, the probability of conversion becomes:

R = $10^{-3}\:(m/\delta m^2)$

For $\delta m^2=10^{-5}eV^2$ and m = 1 eV, the conversion is maximal and the signal easy to find. If such a very favourable situation arises, it becomes possible to measure the absolute value of m by varying the power of the cavity (and consequently R) provided that $\delta m^2$ is known otherwise.

\section{CONCLUSION}

The idea of installing a high quality RF cavity in a high energy neutrino beam is discussed. The case is applied to the K2K beam recently built at KEK. With an existing cavity, it is possible to test radiative decays of neutrinos and to put stringent limits  on the processes (or discover!):

 $\nu_{\mu}\rightarrow\gamma +\nu_{e}$      and      $\nu_{\tau}\rightarrow\gamma +\nu_{\mu}$

The limits reach a level which is of interest for astrophysics and cosmology.

Oscillations only give information on the difference of squared masses. Decays may be the only way to estimate the absolute scale of neutrino masses.
In particular, if the radiative decay of  $\nu_{\mu}$ is responsible for the LSND and SuperKamiokande signals, and assuming that $\delta m^2$ is measured with solar neutrinos, then the stimulated conversion could fix with good accuracy the absolute masses of the participating neutrinos.


\section*{References}
\begin{thebibliography}{99}
\bibitem{fu}Y. Fukuda {\it et al}, \Journal{\PRL}{433}{9}{1998}.

\bibitem{pa}P.B.Pal and L. Wolfenstein, \Journal{\PRD}{25}{768}{1982}.
\bibitem{ch}E.L. Chupp {\it et al}, \Journal{\PRL}{62}{505}{1989}.
\bibitem{kr}D.A. Krakauer {\it et al}, \Journal{\PLB}{252}{177}{1990}.
\bibitem{bo}J. Bouchez {\it et al}, \Journal{\PLB}{207}{217}{1983}.
\bibitem{bi}C. Birnbaum {\it et al}, \Journal{\PLB}{397}{143}{1997}.
\bibitem{do}J.C. D'Olivo, J.F. Nieves and P.B. Pal, \Journal{\PRL}{64}{1088}{1990}.\\
G. Giunti, C.W. Kim and W.P. Lam,  \Journal{\PRD}{43}{164}{1991}.
\bibitem{va}F. Vannucci, Preprint hep-ph/9903487.
\bibitem{ma}S. Matsuki and K. Yamamoto, \Journal{\PLB}{289}{194}{1992}.\\
M.C. Gonzales-Garcia, F. Vannucci and J. Castromonte, \Journal{\PLB}{373}{153}{1996}

\bibitem{oy}Y. Oyama, Preprint hep-ex/9803014.

\end{thebibliography}
\end{document}